# Risk-Aware Financial Forecasting Enhanced by Machine Learning and Intuitionistic Fuzzy Multi-Criteria Decision-Making

Safiye Turgay[1], Serkan Erdoğan[1], Željko Stević[2,3], Orhan Emre Elma[4], Tevfik Eren[4], Zhiyuan Wang[5,*], Mahmut Baydaş[4,*]

[1] Department of Industrial Engineering, Faculty of Engineering, Sakarya University, Sakarya 54050, Türkiye

[2] Faculty of Transport and Traffic Engineering, University of East Sarajevo, 74000 Doboj, Bosnia and Herzegovina

[3] Department of Industrial Management Engineering, Korea University, Seoul 02841, Republic of Korea

[4] Department of Accounting and Financial Management, Faculty of Applied Sciences, Necmettin Erbakan University, Konya 42140, Türkiye.

[5] School of Business, Singapore University of Social Sciences, Singapore 599494, Singapore

\* Corresponding authors: Zhiyuan Wang, Email: zywang@suss.edu.sg; Mahmut Baydaş, Email: mbaydas@erbakan.edu.tr

**Abstract**

In the face of increasing financial uncertainty and market complexity, this study presents a novel risk-aware financial forecasting framework that integrates advanced machine learning techniques with intuitionistic fuzzy multi-criteria decision-making (MCDM). Tailored to the BIST 100 index and validated through a case study of a major defense company in Türkiye, the framework fuses structured financial data, unstructured text data, and macroeconomic indicators to enhance predictive accuracy and robustness. It incorporates a hybrid suite of models, including extreme gradient boosting (XGBoost), long short-term memory (LSTM) network, graph neural network (GNN), to deliver probabilistic forecasts with quantified uncertainty. The empirical results demonstrate high forecasting accuracy, with a net profit mean absolute percentage error (MAPE) of 3.03% and narrow 95% confidence intervals for key financial indicators. The risk-aware analysis indicates a favorable risk-return profile, with a Sharpe ratio of 1.25 and a higher Sortino ratio of 1.80, suggesting relatively low downside volatility and robust performance under market fluctuations. Sensitivity analysis shows that the key financial indicator predictions are highly sensitive to variations of inflation, interest rates, sentiment, and exchange rates. Additionally, using an intuitionistic fuzzy MCDM approach, combining entropy weighting, evaluation based on distance from the average solution (EDAS), and the measurement of alternatives and ranking according to compromise solution (MARCOS) methods, the tabular data learning network (TabNet) outperforms the other models and is identified as the most suitable candidate for deployment. Overall, the findings of this work highlight the importance of integrating advanced machine learning, risk quantification, and fuzzy MCDM methodologies in financial forecasting, particularly in emerging markets.

# 1. Introduction

Today, financial markets are more complex and difficult to predict than ever (Kou & Lu, 2025), in view of increasing geopolitical risks, rapid digitization, and information asymmetry. Such an environment requires financial forecasts that cannot be based on historical data and single-outcome deterministic models alone. There is an emergent need for multilayer forecasting systems which could manage uncertainty, divergence of experts' opinions, and multiple strategic goals of the business effectively. There is a high demand for new-generation forecasting frameworks that are not only of higher accuracy but also aligned with strategic priorities, especially in companies operating in emerging markets. Machine learning models, such as long short-term memory (LSTM) network, time-series Transformer, and graph neural network (GNN), are extensively applied to financial time series forecasting in the existing literature. Multi-criteria decision-making (MCDM) techniques are also being applied to multi-criteria ranking and selection problems extensively (Baydaş et al., 2024). Generally, MCDM is a process concerned with evaluation and selection among alternatives when multiple, often conflicting criteria are considered (Wang et al., 2024). On the other hand, studies incorporating these two approaches into uncertainty sensitivity and group decision-making perspectives are quite scant. Some focus only on the prediction accuracy of artificial intelligence (AI) models, while others stop with rankings based on decision-making criteria. The existing literature gap, therefore, will be systematically attended to in this study by surmounting prior approaches to develop an integrated method of machine learning and multi-criteria decision analytic thinking based on group expert assessment and intuitionistic fuzzy weighting.

Financial statement forecasting is an essential tool supporting investment decisions, corporate risk management, and economic policy. In emerging markets, like the BIST 100 that represents the performance of the 100 largest companies on the Istanbul Stock Exchange in Türkiye, accurately predicting financial statements (e.g., balance sheets, income statements, and cash flow statements) is particularly challenging. These challenges emanate from a few key factors. Severe macroeconomic fluctuations, deep-rooted sectoral dependencies, and firm-specific risks make the realm highly turbulent and intertwined. Traditional approaches to forecasting, like time series regression and econometric models, fall short of appropriately considering nonlinear relationships, intricate data structures, and rapidly changing market dynamics that are characteristic of such a setting.

Therefore, there is increasing demand for AI-driven forecasting models that explicitly account for risks and handle the seamless integration of macroeconomic indicators, investor sentiment, and financial health of firms into their predictive analytics. For example, in Türkiye, macroeconomic uncertainties related to inflation rates, changes in interest rates, and the volatility in exchange rates have an immediate and significant effect on corporate finances. In addition, the financial health of a firm is often dependent on the performance of its peers and supply chain partners, implying that industry and supply chain dependencies can alter either positively or adversely the adaptiveness of a firm.

Compounding these issues is the nature of the data itself: financial data intrinsically possesses a high-dimensional structure with structured components, such as balance sheet figures, but also unstructured ones like news sentiment and earnings call transcripts. The standard machine learning models offer point estimates; however, they mostly fail to provide reliable uncertainty quantification, making them less dependable for the risk-aware financial decision-making required in these markets.

Recent advances in machine learning and AI create unprecedented opportunities for financial statement forecast improvement. Techniques such as LSTM recurrent neural network (RNN), Transformer, GNN, deep reinforcement learning (DRL), tabular data learning network (TabNet), and extreme gradient boosting (XGBoost) have demonstrated exceptional capacity in terms of handling sequential data, complex financial relationships, and text-based sentiment analyses. In addition, Bayesian neural network (BNN) and probabilistic machine learning techniques can help quantify uncertainty; thus, more reliable risk-aware forecasts can be obtained. In this paper, we propose a risk-aware, AI-enhanced financial forecasting framework tailored to the Turkish BIST 100 index. The proposed framework includes k-nearest neighbors (KNN), XGBoost, LSTM, and GNN among several machine learning models to assist the financial forecasting process. It uses a financial bidirectional encoder representations from transformers (FinancialBERT) model to extract macroeconomic indicators and market sentiment. The BNN model is also applied to quantify financial risk and estimate uncertainty within the framework. First, the initial framework was verified using a defense company that is listed on the BIST 100, demonstrating the applicability and effectiveness of the financial performance prediction. In particular, XGBoost was used as the core forecasting model while LSTM captured the temporal dynamics and GNN modeled inter-firm and sectoral dependencies. Then, after validation and sensitivity analysis, the framework is extended by training additional advanced deep learning models for enhanced financial prediction: TabNet, time series transformer, DRL, and GNN. Subsequently, these models were evaluated via an intuitionistic fuzzy MCDM process that considers several performance metrics to determine the most suitable model for a potential deployment within the risk-aware financial forecasting framework.

The novelty of the study lies in the holistic integration of a hybrid machine learning architecture for probabilistic forecasting with a sophisticated intuitionistic fuzzy MCDM process for model selection. To the best of our knowledge, this framework is the first of its kind that can simultaneously: (1) fuse structured financials, unstructured sentiment, and macroeconomic data using models such as XGBoost, LSTM, graph neural network, and BNNs; (2) provide not only point forecasts but also fully quantified confidence intervals and risk scenarios; and (3) utilize an intuitionistic fuzzy Entropy-EDAS-MARCOS approach to transparently rank models based on a comprehensive set of technical, financial, and operational criteria by which expert hesitation is directly integrated into the decision-making process. By bridging these disparate domains of predictive modeling, risk quantification, and group decision-making under uncertainty, this research offers a fresh and practical methodology tailored to the complex volatility of markets like the BIST 100. The research can address the deeper challenge of strategy selection under uncertainty: it offers a structured and transparent process to select a resilient action plan against a variety of future contingencies, turning uncertainty from a paralyzing threat to a measurable dimension of strategic planning. The integration of machine learning and MCDM approaches offers a scalable methodology for dealing with the volatile conditions of emerging markets, where understanding risk is not a secondary concern but a primary determinant of strategic success.

The remaining part of this article is organized as follows: Section 2 reviews the related literature on AI-driven financial forecasting, with a specific focus on applications in emerging markets, such as Türkiye. Section 3 provides an overview of the proposed methodology, including how to develop a risk-aware AI-enhanced fiancial forecasting framework and integrate fuzzy MCDM for model evaluation and selection. Section 4 describes the application of the proposed forecasting framework on a defense company listed on the BIST 100 index. Section 5 presents the experimental results and discussion comprising forecasting performance, sensitivity analyses, and model comparisons through fuzzy MCDM. Finally, Section 6 concludes the paper by summarizing the main contributions, underlining practical implications, and outlining directions for future research.

## 2. Literature Review

The AI applications in financial forecasting have rapidly evolved from conventional statistical models to sophisticated deep learning architectures. However, there is still a significant void in the development of integrated frameworks that are explicitly risk-aware and specialized in the particular volatilities characterizing emerging markets (e.g., BIST 100 in Türkiye). Specifically, relevant research addressing the BIST 100 is at an infancy stage. Though some works apply AI to the Turkish market, most of these studies, such as Mirza et al. (2025) and Altinbas (2025), focus on single model approaches to predict stock prices and also lack a risk-aware perspective. What is significantly absent is the usage of hybrid frameworks that integrate diverse data sources, such as financial, macroeconomic, and sentiment aspects, employ probabilistic models with quantification of uncertainty, and use advanced MCDM methods for comprehensive model evaluation. This section reviews the literature across these key thematic areas, leading to a summary table that demarcates the progress and limitations of existing research.

During the last years, AI-based forecasting models have been able to achieve extraordinary success in capturing the complex dynamics that are inherent in financial time series forecasting, as presented by Khattak et al. (2023) and Olubusola et al. (2024), and in promoting financial risk analysis, as indicated by Kou et al. (2019) and Shen et al. (2021). The advanced models of LSTM networks, Transformers, TabNet, and GNNs exhibited quite robust capabilities in reducing forecast error by efficiently modeling nonlinear relationships and temporal dependencies, as mentioned by Li & Law (2024). In this section, an intensive review is provided of the existing literature on AI-driven methods for the forecast of financial statements, with specific attention being devoted to applications related to BIST 100 companies. This review covers major methodologies, prevailing challenges, and recent advances that have been made in highlighting both the strengths of state-of-the-art approaches and critical gaps that this study tries to bridge with a proposed hybrid risk-aware framework.

### 2.1. Evolution from Traditional to AI-Driven Forecasting

Traditional models include autoregressive integrated moving average (ARIMA) and exponential smoothing, which have been widely adopted for financial time series forecasting. According to Rubio et al. (2021), the significant strength of these traditional models is associated with modeling linear dependencies; however, they always suffer from the non-linear chaotic nature of financial markets. The advent of machine learning models such as XGBoost (Ali et al., 2023) perform excellently on structured tabular data due to robustness and feature importance analysis. For sequential data, LSTM networks is popular for capturing temporal patterns, overcoming the vanishing gradient problem of traditional RNNs. More recently, Transformer models have been adapted for time series analyses, making use of self-attention mechanisms in an attempt to capture long-range dependencies more effectively than LSTMs.

However, most of these methods always face challenges in detecting nonlinear patterns/structures in the data, as noted by Muthamizharasan & Ponnusamy (2024), and complex interrelations among financial variables. Current studies have integrated AI models, especially LSTM networks, which are excellent in learning sequential patterns and trends that occur with time using the gating mechanisms (Wang et al., 2023). This makes them relevant to any time series forecasting, according to Abotaleb & Dutta (2024). Another newly developed model involves the Transformer model, characterized by their attention mechanisms that have been helpful in the capturing of long-term dependencies in data, as noted by Thundiyil et al.

(2023). XGBoost is also a tree-based ensemble learning algorithm that has gained wide acceptance because of its high predictive accuracy, handling of missing values, and robustness in modeling structured financial data, as confirmed by Ali et al. (2023). GNN is an up-and-coming popular deep learning technique in financial forecasting because of its ability to model relationships among different entities, such as companies and stakeholders, and capture complicated interactions within financial ecosystems, as reviewed by Wang et al. (2021).

## 2.2. Integrating Alternative Data with Market Structure

This has been the basis of a critical development in forecasting: moving beyond purely quantitative data. Macroeconomic indicators of inflation and interest rates, for example, are now seen as integral in setting a wider context within which firm performance is assessed (Khan et al., 2024). More recently, natural language processing (NLP) methods, particularly models like BERT, perform sentiment analysis from news and social media to provide a qualitative angle to forecasts (Yenduri et al., 2024; Yadav, 2024). In modeling the linked nature of markets, GNNs have been developed that can model relational structures, thereby identifying systemic risks from the links between companies and sectors (Wang et al., 2021).

## 2.3. Shift to Quantification of Risk and Uncertainty

Although valuable, point forecasts are not sufficient for risk-sensitive decision-making. This has sparked an interest in probabilistic forecasting. Bayesian methods and Monte Carlo simulations are used to quantify uncertainty, providing confidence intervals and predictive distributions (Moolchandani, 2023; Zheng, 2024). BNNs formally model epistemic (model) and aleatoric (data) uncertainty, thus providing a principled framework for risk-aware predictions.

These methods provide further insight into the uncertainty of future financial outcomes, which is critical for risk management and decision-making. Various studies have established the efficacy of these probabilistic techniques in various financial applications, such as revenue forecasting, stock price predictions, and financial performance measures. These methods allow for better decision-making by incorporating uncertainty into the analysis, especially in highly volatile environments like emerging markets.

While the research in the application of AI techniques is still in the early stages in the Turkish market, some development has been made in adapting the forecasting models to the peculiar characteristics of the Turkish economy. The Turkish market, especially the BIST 100 companies, faces certain challenges such as currency fluctuations, inflation, and geopolitical risks that require specialized forecasting models (Altinbas, 2025). A few studies have analyzed the stock market and financial forecasting in the context of the Turkish market, where models that could adapt to the local and global economic influences were called for (Mirza et al., 2025). However, no integration of hybrid AI models, such as the integration between XGBoost, LSTM, and GNN for forecasting financial statements in the Turkish market, has been considered. This opens up a very valuable opportunity for the proposed framework to contribute meaningfully.

Despite recent advances in AI-driven financial forecasting, substantial gaps persist in the literature, with very few attempts to combine disparate data sources, including macroeconomic indicators, market sentiment, and firm-specific financial statements, into a single unified, risk-aware model. Thus, although promising, most of the works are still lacking in terms of truly representing the uncertainty and risk in a full-fledged manner. This paper proposes an integrated AI-boosted hybrid framework that combines KNN, FinancialBERT, XGBoost, LSTM, GNN, and BNN to yield more comprehensive and risk-aware financial predictions. The proposed framework has been tested on a large defense company listed in the BIST 100 index and yields practical insights regarding the use of machine learning in enhancing forecasting and risk assessment within the Turkish market.

## 2.4. Multi-Criteria Decision-Making in Model Selection

The selection of the most appropriate model becomes a complex decision problem as the arsenal of available ML models grows. MCDM methods provide a structured approach in which alternatives are evaluated against multiple, often conflicting, criteria (Ding et al., 2025). Fuzzy extensions of MCDM have been developed, especially intuitionistic fuzzy sets, to handle the inherent subjectivity and imprecision in expert judgments. Methods such as EDAS and MARCOS have been successfully applied in various domains to provide robust rankings of alternatives (Wang & Rangaiah, 2025).

After the validity and applicability of this core framework are verified, the study is extended by incorporating other advanced deep learning models (including TabNet, time-series Transformer, DRL, and GNN), which are independently trained based on the preprocessed dataset. Further, these models are evaluated using an intuitionistic fuzzy MCDM process founded on the fuzzy group decision-making (Li et al., 2021), considering a wide range of performance metrics to identify the most suitable model to be used for possible implementation within the proposed risk-aware financial forecasting framework. Fuzzy MCDM represents an advanced decision-support approach, extending the traditional MCDM by considering fuzzy extensions (Zulqarnain et al., 2021; Kou et al., 2021), while in intuitionistic fuzzy MCDM, each evaluation represents intuitionistic fuzzy numbers (IFNs), which are defined by a membership degree, a non-membership degree, and a hesitation margin capturing the indecisiveness or incomplete information of the decision maker. This further developed structure allows the modeling of expert opinions in a way that is more nuanced and realistic (Stanković et al., 2020), particularly in complex or uncertain environments, such as financial forecasting. Intuitionistic fuzzy MCDM is therefore well tailored for aggregating group judgments, as it allows the accommodation of diverse perspectives while accounting for vagueness in the evaluation of criteria. In this context, it will allow for the flexible yet rigorous comparison of alternatives, combining fuzzy logic, modeling uncertainty, and multi-criteria analysis within a single integrated framework. A summary of the literature review is presented in Table 1.

Table 1. Summary of literature survey on AI-driven financial forecasting.

| Authors (Year) | Methodology | Key Findings / Pros | Cons / Limitations |
|---|---|---|---|
| **Traditional & Single-Model Approaches** | | | |
| Rubio et al. (2021) | ARIMA, Exponential Smoothing | Effective for linear trend decomposition and short-term forecasting of EBITDA | Does not catch non-linear patterns and external shocks; quantification of uncertainty not done |
| Ali et al. (2023) | XGBoost for financial fraud detection | High predictive accuracy, interpretability for structured financial data | Focus is on the classification of fraud, not time series forecasting or risk assessment. |
| Yu et al. (2019) | LSTM networks | A comprehensive review confirms its superiority to RNNs for capturing long-term temporal dependencies. | Can be computationally expensive, results are usually point forecasts without confidence intervals. |
| **Advanced Architectures & Hybrid Data** | | | |
| Thundiyil et al. (2023) | Transformer models for time-series | Attention mechanisms efficiently grasp long-range dependencies in sequential data. | Require large amounts of data; can be less efficient compared to LSTMs on smaller datasets. |
| Wang et al. (2021) | Graph Neural Networks (GNN) | It models both interfirm relationships and sectoral dependencies effectively for better forecasts. | Complexity in graph construction; computational overhead. |
| Yenduri et al. (2024) | BERT & GPT for NLP | High accuracy with respect to extracting contextual sentiment analysis from financial text | Primarily focused on sentiment extraction, not integrated into a full forecasting pipeline |
| **Risk & Uncertainty Focus** | | | |
| Zheng (2024) | Bayesian Networks | These are useful in project financing risk assessment because the modeling allows for conditional dependencies. | Often applied purely for the purpose of risk analysis, and not linked to predictive forecasting. |
| Moolchandani (2023) | Monte Carlo Simulations | Strong in quantifying uncertainty and generating probability distributions of outcomes. | Often utilized in isolation, not integrated with modern deep learning models for end-to-end forecasting. |
| **MCDM for Model Evaluation** | | | |
| Stević et al. (2020) | MARCOS MCDM Method | Provides a stable and efficient compromise solution for the selection of suppliers. | Applied in supply chain, not for evaluating and ranking financial forecasting models. |
| Kou et al. (2021) | Fuzzy MCDM for FinTech | Effectively manages vagueness of expert judgments for the selection of investments. | Focus is on investment decisions, not on the selection of underlying forecasting models. |
| **Research on Turkish Markets** | | | |
| Altinbas (2025) | Event Study Analysis | Identifies specific geopolitical and economic events driving volatility in BIST. | Descriptive analysis of the causes of volatility, but not a predictive forecasting model. |
| Mirza et al. (2025) | CNN-RNN Hybrid | A novel architecture which outperforms the benchmark techniques in BIST for stock price forecasting. | Focuses on price prediction only; financial statement forecasting and risk quantification are ignored. |

## 3. Methodology

### 3.1. Overview of Risk-Aware Financial Forecasting Framework and Components

This research methodology encompasses several key phases: data collection, feature engineering, model selection, risk quantification, and performance evaluation. Each phase is important in making financial forecasts robust and reliable. Data collection starts with the gathering of structured and unstructured financial data from various sources. On the one hand, structured data is obtained by collecting company financial reports, such as quarterly and annual balance sheets from the BIST 100 firms, through the Public Disclosure Platform (KAP, kap.org.tr/en, for the Turkish capital markets). The structured data also consists of macroeconomic indicators-including inflation, interest rates, exchange rates, and gross domestic product (GDP) growth obtained from the Turkish Statistical Institute (TÜİK) and the Central Bank of the Republic of Türkiye (TCMB). To capture the dynamics of market conditions, it also considers the collection of historical stock prices, trading volumes, and volatility measures from BIST. On the other hand, unstructured data, which includes financial news, earnings call transcripts, and analyst reports, is preprocessed using NLP techniques to extract sentiment and depict qualitative market insights.

The preprocessing phase involves handling missing data, feature scaling, and text preparation. Missing values are imputed using advanced techniques such as KNN and expectation-maximization. For feature scaling, both min-max scaling and Z-score normalization are applied to ensure consistency across the dataset. Textual data undergoes preprocessing steps including tokenization, stopword removal, and vectorization using methods like term frequency-inverse document frequency (TF-IDF) and word embeddings such as Word2Vec and BERT to extract meaningful features. Additionally, time-series features are engineered by creating lag variables, computing rolling averages, and applying exponential smoothing to capture the temporal dynamics within the financial data.

Figure 1 illustrates a hybrid artificial intelligence technique, which integrates deep learning, probabilistic forecasting, and graph-based methods for model selection. In brief, LSTM networks can be leveraged to capture temporal dependencies in sequential data arising from structured financial data. Gradient boosting-based models, such as XGBoost, have been incorporated for accurate regression forecasting along with feature importance analysis. Unstructured data like financial news and earnings calls is assessed using BERT-based sentiment analysis. GNNs are adapted to learn sectoral dependencies and firm linkages. Risk-aware forecasting is one of the important approaches within the methodology. BNNs can quantify uncertainty in financial predictions by obtaining probability distributions for forecasts instead of point estimates. Monte Carlo Dropout is used to estimate confidence intervals, providing an understanding of the uncertainty in predictions. Quantile regression has been made use of to provide high-range and low-range forecasts, enabling the consideration of worst-case and best-case financial scenarios.

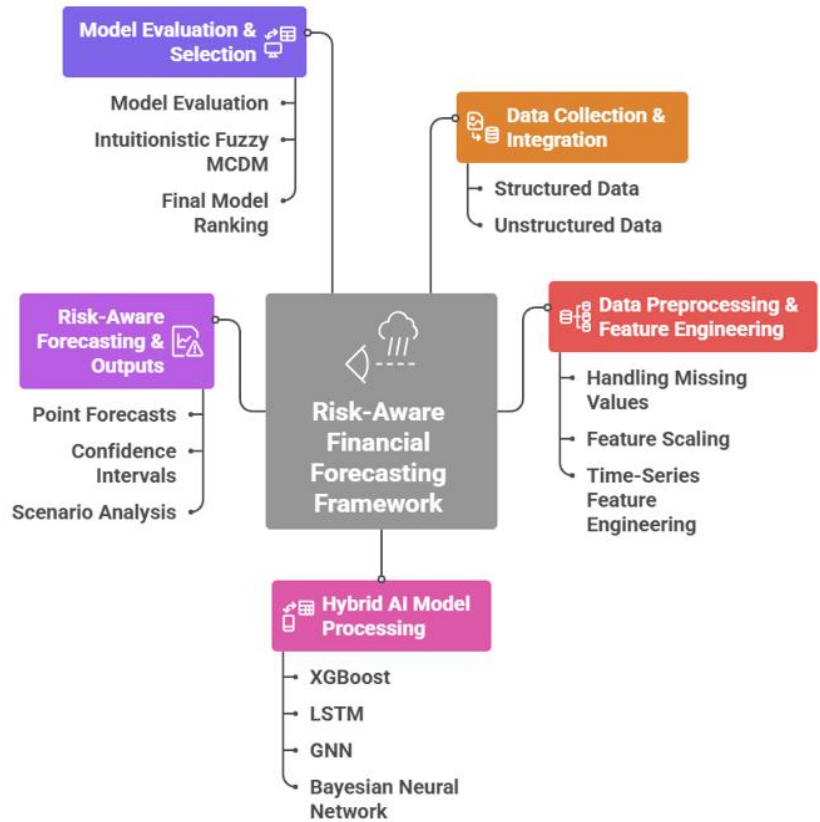

Figure 1. Components for the risk-aware financial forecasting framework.

As aforementioned, the major contribution of the developed model lies in a deliberate and synergistic design that consciously and coherently combines XGBoost, LSTM, GNN, BNN, and intuitionistic fuzzy MCDM to handle the highly complex nature of risk-aware financial forecasting in an interconnected and uncertain environment. Each component undertakes the responsibility of solving a certain and crucial sub-problem that cannot be comprehensively addressed by conventional mono-models. For instance, by nature, financial data, such as balance sheets and ratios, is structured and tabular, and then XGBoost is the state-of-the-art ensemble method providing very accurate and robust point estimates from this type of data. It sets a performance metric by baselining robustly and interpretably, treating nonlinear relationships within the static snapshots much more efficiently than most deep learning models. On the other hand, LSTMs are specifically designed for modeling sequential dependencies and hence predict future outcomes by learning patterns in historical data, for instance, company performance over the last eight quarters. While XGBoost can make use of delayed features, the internal memory cells of LSTM provide the possibility of capturing complex, long-term temporal trends and cycles, which are important for accurate forecasting. Company performance does not exist in a vacuum but is interlinked with peers, competitors, and partners within the same sector (e.g., BIST 100 firms). Traditional models do not take these relational structures into account. GNNs model this interconnectedness explicitly, thereby enabling the framework to model contagion risk and sectoral trend capture. If a key partner stock goes down, for instance, the GNN can factor this into the forecast for the more holistic, system-focused forecast. However, point forecasts are not sufficient for risk-sensitive decision-making. BNNs essentially answer the question "How reliable is the model?" through placing probability distributions over model parameters. The framework can thus allow differentiating aleatoric uncertainty (i.e., the inherent noise in the data) from epistemic uncertainty (i.e., model uncertainty because of lack of data). Instead of a single output value, we get a predictive distribution enabling the calculation of confidence intervals and VaR (Value at Risk), which is an indispensable element of risk management.

Additionally, the need to develop an intuitionistic fuzzy MCDM (i.e., Entropy-EDAS-MARCOS) strategy arises because various advanced models produce different predictions, necessitating a multi-faceted and rational selection process. The MCDM technique evaluates models across multiple criteria, including computational efficiency, prediction accuracy, Sharpe ratio, and others. In essence, the logic of our proposed framework is both sequential and complementary: XGBoost and LSTM make robust baseline predictions from different data perspectives (static and temporal). GNN enriches these with contextual and relational intelligence. BNN wraps them in a probabilistic layer that quantifies the uncertainty of their aggregated predictions. Finally, fuzzy MCDM offers a rational mechanism for selecting the most suitable model for deployment, not on the basis of a single technical metric but through a comprehensive set of strategic criteria. It is this hybrid design that allows for a true shift from pure prediction to actionable, risk-aware decision support.

### 3.2. Theoretical Basis and Formal Definitions

A basic hypothesis of the framework is that any robust financial forecasting needs to incorporate nonlinear relationships, exogenous risk factors, and predictive uncertainty. Let $\mathbf{X}_t$ denote the structured financial data at time $t$, such as features extracted from the balance sheet, income statement, or financial ratios. Let $\mathbf{S}_t$ represent the unstructured data, including textual signals like sentiment scores derived from news articles, earnings call transcripts, or social media. The model's forecast for time $t+1$, denoted by $\hat{Y}_{t+1}$, is expressed as:

$$\hat{Y}_{t+1} = f(\mathbf{X}_t, \mathbf{S}_t, \boldsymbol{\theta}) + \varepsilon_t \qquad (1)$$

where $f(\cdot)$ is the predictive function, learned from the data using machine learning methods such as XGBoost, LSTM, or Transformer-based architectures; $\boldsymbol{\theta}$ denotes the model parameters, which are optimized during training; and $\varepsilon_t$ is the residual error term, capturing noise and imperfections in the model's prediction. It accounts for both inherent randomness in the data (aleatoric uncertainty) and model limitations (epistemic uncertainty).

A risk-aware forecast provides a full probability distribution for the target variable. More formally, the forecast is given by:

$$\hat{Y}_{t+1} \sim \mathcal{N}(\mu_{t+1}, \sigma_{t+1}^2) \qquad (2)$$

where $\mu_{t+1}$ is the mean prediction, and $\sigma_{t+1}$ is the standard deviation which quantifies the level of uncertainty (risk) associated with the forecast.

In a BNN, the model parameters $\boldsymbol{\theta}$ are treated as random variables with a prior distribution that gets updated to posteriors given observed data. Finally, its predictive distribution integrates over all parameters:

$$p(\hat{Y}_{t+1} \mid \mathbf{X}_t, \mathbf{S}_t) = \int p(\hat{Y}_{t+1} \mid \mathbf{X}_t, \mathbf{S}_t, \boldsymbol{\theta}) p(\boldsymbol{\theta} \mid \mathbf{X}_t, \mathbf{S}_t) \, d\boldsymbol{\theta} \qquad (3)$$

This naturally generates prediction intervals in which there is 95% probability that the true value lies within $\hat{Y}_{t+1} \pm 1.96\sigma_{t+1}$. This provides a mathematically sound measure of forecast confidence.

$$p(Y_{t+1} \in [\hat{Y}_{t+1} - 1.96\sigma_{t+1},\ \hat{Y}_{t+1} + 1.96\sigma_{t+1}]) = 0.95 \qquad (4)$$

### 3.3. Machine Learning Development, Optimization, and Dataset

Machine learning has become widely applied in diverse fields to enable predictive modeling based on large datasets (Wang et al., 2022). Its popularity stems from the capacity to uncover complex patterns and improve performance over time. This section provides a detailed insight into the development pipeline, strategies of optimization, and how the dataset is constructed under this risk-aware forecasting framework. The developed methodology targeted model robustness, reproducibility, and fairness across cross-model evaluations.

#### 3.3.1. Data Collection and Preprocessing

This dataset integrates structured, unstructured, and macroeconomic data from 2018 to 2024, which can provide a multidimensional understanding of the defense company and its financial environment. In the structured portion, 5 years of quarterly financial statements (i.e., 20 periods) for the target defense company and key peers in the BIST 100 are included, with more than 150 features, such as debt-to-equity and current ratio for financial ratios, raw indicators of total assets and revenues, and a wide range of growth metrics. In the unstructured dataset, more than 15,000 financial news articles and 120 earnings-call transcripts related to the Turkish defense sector have been gathered from KAP and major national news agencies. Quarterly macroeconomic indicators of CPI inflation, BIST 100 interest rates, USD/TRY exchange rate, and GDP growth rates have been collected from TÜİK and TCMB to put firm-level dynamics into perspective. A multistage preprocessing pipeline was developed to guarantee data quality and analytic consistency. Missing financial data, less than 5% of all entries, was imputed using the KNN approach with k = 5, since it is superior to other mean-based or median-based approaches in terms of maintaining multivariate relationships. All numerical variables were standardized using Z-score normalization to obtain zero mean and unit variance, which is crucial for the stable and efficient training of neural network architectures. Regarding text data, documents were cleaned via stopword removal and tokenization, after which a fine-tuned FinancialBERT model, pre-trained on Turkish financial news, generated sentiment scores ranging from –1 (strongly negative) to +1 (strongly positive). Last but not least, for the GNN framework, the dynamic temporal graph for every quarter was designed: nodes corresponded to firms, and edges were weighted using the correlation of stock returns over the previous four quarters, adjusted by the categorical relationship between sectors. Including the integration of structured and unstructured data with a complex relational structure among observations, this provides a comprehensive, high-fidelity forecasting environment.

#### 3.3.2. Model Architectures and Hyperparameter Optimization

All predictive models have been developed and fine-tuned using the same consistent and rigorous methodology. Implemented in Python using PyTorch and Scikit-learn libraries, models were trained and evaluated on a chronologically ordered dataset. Data ranging from 2018-2022 consisted of approximately 80% of the training set, while data from 2023-2024 consisted of the remaining approximate 20% used for testing. To make the model robust against temporal dependencies, a 5-fold time-series cross-validation strategy was adopted only for the training portion. Hyperparameter optimization was performed via Bayesian Optimization using a Tree-structured Parzen Estimator (TPE) through the Optuna framework. Each model was optimized for 100 trials, where the objective function was to minimize mean absolute error (MAE) in the validation folds. This is a systematic tuning targeting key hyperparameters for each architecture to ensure that all models have their best possible configuration under consistent conditions when their performances are compared in Table 2.

Table 2. Key hyperparameters tuned for each model.

| Model | Key Hyperparameters Tuned | Optimal Ranges Found |
| --- | --- | --- |
| **XGBoost** | n_estimators (500-2000), max_depth (6-12), learning_rate (0.01-0.1), subsample (0.7-0.9), colsample_bytree (0.7-0.9) | n_estimators: 1150, max_depth: 9, learning_rate: 0.05 |
| **LSTM** | Layers (1-3), Units/Layer (32-128), Dropout Rate (0.1-0.4), Learning Rate (1e-4 to 1e-2), Sequence Length (4-8 quarters) | 2 Layers, 64 Units, 0.2 Dropout, Seq. Len: 6 |
| **GNN** | Graph Conv Layers (2-4), Hidden Channels (16-64), Learning Rate (1e-4 to 1e-2), Pooling Method (mean, max, attention) | 3 Layers, 32 Channels, Attention Pooling |
| **TabNet** | n_d/n_a (8-64), n_steps (3-8), gamma (1.2-1.6), momentum (0.8-0.99) | n_d=32, n_a=32, n_steps=5, gamma=1.4 |
| **Time-Series Transformer** | Encoder Layers (2-4), Attention Heads (4-8), Feed-Forward Dimension (128-512), Learning Rate (1e-4 to 1e-3) | 3 Encoder Layers, 8 Heads, 256 FF Dimension |
| **BNN (for Uncertainty)** | Layers (2-3), Units (32-64), Dropout Rate (for MC Dropout) (0.1-0.4), Prior Distribution (Gaussian, Scale) | 2 Layers, 48 Units, 0.25 Dropout |

### 3.3.3. Training and Implementation

The training and implementation pipeline was planned in a way that it would give strong performance to the model with a view to limiting overfitting and improving generalization. For the XGBoost model, early stopping with a patience of 50 rounds based on validation loss was done to allow its halt before overfitting. The neural network models, which include LSTM, Transformer, GNN, and BNN, were all trained using the AdamW optimizer with a weight decay of 1e-4 to enhance regularization. A learning rate scheduler, ReduceLROnPlateau, was used, which adaptively reduced the learning rate upon the stagnation of validation performance. Each of these models was trained with a batch size of 32 and early stopping to prevent overfitting with a patience of 15 epochs. Uncertainty quantification was performed in BNNs; during inference, Monte Carlo Dropout was utilized by executing 100 stochastic forward passes to get each prediction. The mean of these predictions was taken as the final output, while the standard deviation over samples formed the basis of constructing the 95% confidence interval. Finally, the DRL module has been implemented by using a Deep Q-Network (DQN) agent, which was trained with an experience replay buffer.

Traditional models, like ARIMA, generate point forecasts using only historical time series data. Nevertheless, they generally do not consider uncertainty or external factors such as macroeconomic volatility, policy changes, or market sentiment that may highly impact a firm's future performance. In contrast, the proposed approach uses Bayesian methods and Monte Carlo sampling in order to model uncertainty explicitly. This enhances the reliability of the forecasts by providing not only expected values but also confidence intervals and predictive distributions, making the outputs more robust and actionable for risk-sensitive decision-making. Besides, GNN further enhances predictive performance by modeling inter-firm and inter-sectoral dependencies that are usually ignored by traditional models. If the keystone firm in a supply chain or the leading firm in an industry is in distress, for example, this can cascade into affecting the viability of other firms connected to it. The GNN is designed to model these kinds of relational structures, thereby facilitating the capture of collective risk exposure in the proposed method. This results in even more realistic and informed financial forecasts that account for systemic effects in economic networks.

### 3.4. Intuitionistic Fuzzy Entropy-EDAS-MARCOS Approach for Fuzzy MCDM

In this study, the ranking and selection of the most appropriate deep learning model for the integration into the risk-aware financial forecasting framework are conducted using the intuitionistic fuzzy Entropy-EDAS-MARCOS approach, which seamlessly integrate the entropy weighting method, EDAS (Keshavarz Ghorabaee et al., 2015), and MARCOS MCDM methods (Stević et al., 2020). The intuitionistic fuzzy theory is an extension of classical fuzzy set theory by using an additional dimension of uncertainty that is called hesitation. In an IFN, every assessment is portrayed as a pair $\tilde{x}_{ij} = (\mu_{ij}, v_{ij})$, where $\mu_{ij}$ represents the degree of membership (i.e., to what extent an alternative satisfies a certain criterion) and $v_{ij}$ represents the non-membership degree (i.e., to what extent it does not). These values need to satisfy this condition: $0 \leq \mu_{ij} + v_{ij} \leq 1$.. The hesitation degree, given as $\pi_{ij} = 1 - \mu_{ij} - v_{ij}$, shows the residual uncertainty or indecisiveness in the evaluation, considering incomplete and imprecise information. This three-part structure allows intuitionistic fuzzy theory to model expert judgments more flexibly and accurately in complex decision scenarios.

Let $A_i$ denote the $i$-th alternative, $i = \{1,2,...,m\}$; $C_j$ denote the $j$-th criterion, $j = \{1,2,...,n\}$; $\tilde{x}_{ij} = (\mu_{ij}, v_{ij})$ be the IFN representing the evaluation of $A_i$ with respect to $C_j$; $w_j$ be the normalized weight of criterion $C_j$, and $\sum_{j=1}^{n} w_j = 1$; $\pi_{ij} = 1 - \mu_{ij} - v_{ij}$ denote the hesitation degree.

The next step is to determine objective criterion weights using one of the fuzzy variants of the entropy weighting methods. By this method, the informational contribution of each criterion for all alternatives is quantified. For every criterion, the corresponding weight $w_j$ is obtained:

$$w_j = \frac{1 - E_j}{\sum_{j=1}^{n}(1 - E_j)} \qquad (5)$$

where $E_j$ represents the entropy of the criterion. With the weights obtained, the core concept of EDAS is applied. For each criterion $C_j$, compute the average membership and non-membership values across all alternatives:

$$\mu_j^{\text{avg}} = \frac{1}{m}\sum_{i=1}^{m}\mu_{ij}, \qquad v_j^{\text{avg}} = \frac{1}{m}\sum_{i=1}^{m}v_{ij} \qquad (6)$$

This gives the average intuitionistic fuzzy solution $\tilde{x}_j^{\text{avg}} = (\mu_j^{\text{avg}}, v_j^{\text{avg}})$, which acts as a reference point.

Then, the positive distance from average (PDA) and the negative distance from average (NDA) are calculated for every alternative and criterion:

For benefit-type criteria:

$$PDA_{ij} = \max\left(0, \mu_{ij} - \mu_j^{\text{avg}}\right), \qquad NDA_{ij} = \max\left(0, \mu_j^{\text{avg}} - \mu_{ij}\right) \qquad (7)$$

For cost-type criteria:

$$PDA_{ij} = \max\left(0, \mu_j^{\text{avg}} - \mu_{ij}\right), \qquad NDA_{ij} = \max\left(0, \mu_{ij} - \mu_j^{\text{avg}}\right) \qquad (8)$$

Here, the PDA represents the amount that alternative exceeds the average and the NDA measures the amount it falls below the average. The total weighted PDA and NDA then for each alternative is computed as:

$$PDAS_i = \sum_{j=1}^{n} w_j\, PDA_{ij}, \quad NDAS_i = \sum_{j=1}^{n} w_j\, NDA_{ij} \qquad (9)$$

To transform the values into one simple performance indicator, the adjusted appraisal score ($S_i$) is defined as:

$$S_i = \frac{PDAS_i}{PDAS_i + NDAS_i} \qquad (10)$$

The final step applies the core concept of the MARCOS method by Baydaş et al. (2025), to define the ranking of alternatives. According to the previously calculated appraisal scores $S_i$, the following steps are done: Then identify the best solution and anti-ideal solution as:

$$I^* = \max(S_i), \qquad I^- = \min(S_i) \qquad (11)$$

Calculate utility function value $U_i$ for each alternative by the formula:

$$K_i^+ = \frac{S_i}{I^*}, \qquad K_i^- = \frac{I^-}{S_i} \qquad (12)$$

$$U_i = \frac{K_i^+ + (1 - K_i^-)}{2} \qquad (13)$$

The utility function, which incorporates compromise-based reasoning, measures each alternative against both the best and worst performance. The alternative that has the highest $U_i$ is viewed as most favorable and therefore ranked first.

## 4. Application

In this section, we demonstrate how the risk-aware financial statement forecasting framework can be applied to a defense company from the BIST 100 index in Türkiye. The purpose of this case study is to address how a hybrid AI model integrates financial data and sentiment analysis to predict the future financial outcome of the company while considering its risks. This defense company is selected due to its significant market share, its recent government contracts, and its sensitivity to the domestic and international economic situation, making it particularly suitable for risk-aware financial forecasting. To predict the financial statements of the firm, we need both structured and unstructured data. The structured financial data includes quarterly and annual balance sheets, income statements, and cash flow statements from the past five years. Unstructured data is comprised of textual information from the firm's quarterly earnings calls, press releases, and news articles covering defense contracts and geopolitical events. Sentiment scores are then calculated from these texts using FinancialBERT. The macroeconomic data, such as interest rates, inflation rates, exchange rates obtained from TÜİK and TCMB, are also added. We also consider market sentiment and sector dependencies by incorporating knowledge of the performance in the defense sector, the stock correlations of firms in the sector, and global defense spending.

The forecasting process is organized into several steps. Data preprocessing is emphasized at the first stage. Missing values are handled in structured data using the KNN imputation technique, while normalization of financial features is performed, such as total assets, liabilities, and equity. The preparation of textual data for sentiment analysis is performed, which, after processing, generates sentiment scores using FinancialBERT. In addition, external macroeconomic data is standardized, including inflation and interest rates. We then train the models by using XGBoost to predict key financial indicators, including revenue, net profit, and total assets. Also, we train an LSTM time-series model on historical stock prices and financial data of the company to model temporal dependencies for future stock price and financial trend predictions. A GNN is also used in order to model sectoral interdependencies by taking into account the interdependencies between the company and other companies in the defense sector.

The next step is dedicated to uncertainty quantification. We estimate uncertainty in financial predictions by using a BNN, leveraging Monte Carlo Dropout for an approximate calculation of the posterior distribution over model weights. We apply quantile regression for both the 10th percentile (worst-case) and the 90th percentile (best-case) scenarios of the financial performance of the company. Finally, we get to the forecasting results for short term (1-2 quarters), mid-term (1-2 years), and also the risk-aware scenarios.

This case study then outlines the expected short-term (2024-Q2) financial performance of the company, generated by the risk-aware financial statement forecasting framework, as depicted in Table 3. These estimates are done via training AI models on historical data, considering sentiment analysis, macroeconomic indicators, external risks, and sectoral interdependencies. It also covers confidence intervals and risk-aware scenarios, that is, best-case, base-case, and worst-case scenarios, to provide a wider range of possible outcomes that decision makers can take into consideration when preparing for market conditions. This includes geopolitical factors and sentiment analyses to determine how influential external risks may be on the predicted financial performance, just by the differences between the best-case and worst-case scenarios.

Table 3. Short-term (2024-Q2) forecast values by risk-aware forecasting model.

| Category | Metric | Value (in million TRY) |
| --- | --- | --- |
| **Short-Term (2024-Q2) Forecast by XGBoost** | Total Assets | 15,150 |
| | Net Profit | 1,600 |
| | Equity | 8,100 |
| **Risk Quantification by BNN** | Total Assets (95% Confidence Interval) | 15,150 ± 150 |
| | Net Profit (95% Confidence Interval) | 1,600 ± 100 |
| | Equity (95% Confidence Interval) | 8,100 ± 80 |
| | Best-case Scenario (Net Profit) | 1,750 |
| | Worst-case Scenario (Net Profit) | 1,400 |
| **Risk-Aware Scenarios** | Best-case Scenario - Total Assets | 15,250 |
| | Best-case Scenario - Net Profit | 1,750 |
| | Best-case Scenario - Equity | 8,150 |
| | Base-case Scenario - Total Assets | 15,150 |
| | Base-case Scenario - Net Profit | 1,600 |
| | Base-case Scenario - Equity | 8,100 |
| | Worst-case Scenario - Total Assets | 15,050 |
| | Worst-case Scenario - Net Profit | 1,400 |
| | Worst-case Scenario - Equity | 8,000 |

We perform a numerical comparison of the predictions made by the risk-aware financial statement forecasting model, trained on the company's historical financial data, against the actual financial results for 2024-Q2. The model is then evaluated based on its performance in predicting the key financial metrics according to a comprehensive set of performance metrics: accuracy, uncertainty quantification, and risk-adjusted performance. The accuracy of the forecast is ensured through metrics, such as MAE and mean absolute percentage error (MAPE), while risk is quantified through appropriate metrics such as the width of confidence interval and VaR. Risk-adjusted performance is further analyzed with the Sharpe Ratio, the Sortino Ratio, and scenario analyses considering both worst-case and best-case conditions. A comparison of predicted and actual values is shown in Table 4.

Table 4. Comparison of XGBoost-predicted and actual financial metrics for 2024-Q2 with error measures (MAE and MAPE).

|  | Predicted Values by XGBoost (in million TRY) | Actual Values (in million TRY) | MAE (in million TRY) | MAPE (%) |
|---|---|---|---|---|
| **Total Assets** | 15,150 | 15,200 | 50 | 0.33% |
| **Net Profit** | 1,600 | 1,650 | 50 | 3.03% |
| **Equity** | 8,100 | 8,120 | 20 | 0.25% |

The MAE for total assets is 50 million TRY, which is reasonable given the scale of a defense company's balance sheet. The MAPE of net profit is 3.03%, which means a good level of forecasting accuracy, though greater sensitivity may be needed in highly volatile scenarios.

Moreover, the risk quantification component in Table 3 expresses confidence intervals around the forecasts, permitting an evaluation of forecast uncertainty. The 95% confidence intervals of total assets, net profit, and equity are relatively narrow, indicating that the model's forecasts are subject to a low degree of uncertainty under current market conditions. Confidence intervals provide insights into possible fluctuations in finance, which have shown that the financial performance of the company remains relatively stable under normal market conditions. For example, at 95% confidence, VaR for net profit is 150 million TRY, which means there is only a 5% chance the company's net profit may fall more than this amount. After that, we look into the Sharpe Ratio and Sortino Ratio, assessing the risk-adjusted return relative to the company's projected financials. Both of these metrics have been considered fundamentally essential to return evaluation with respect to their associated risks. Specificlialy,the Sharpe Ratio captures total volatility, while the Sortino Ratio focuses on downside volatility. The Sharpe Ratio calculated equals 1.25, whereas the calculated value of the Sortino Ratio is 1.80. A Sortino Ratio that is higher means the financial performance of the company is accompanied by relatively low downside volatility. This implies a good risk-return profile, which means that the company is very likely to sustain good performance even in negative market conditions.

To rigorously quantify the value added by the proposed hybrid ML framework, its forecasting performance is benchmarked against that of a traditional statistical model, namely the ARIMA model. This ARIMA model is trained on the same historical financial data from 2018 to 2022, and its hyperparameters (p, d, q) have been optimized by standard time-series techniques. The out-of-sample forecast for 2024-Q2 is shown in Table 5.

Table 5. Comparisons with the traditional ARIMA Model (2024-Q2).

| Metric | ARIMA (Benchmark) | XGBoost (Core ML Model) | Improvement vs. ARIMA | Full Hybrid Framework (Best Model) |
|---|---|---|---|---|
| Net Profit MAPE (%) | 5.82% | 3.03% | -47.9% | 2.91% (TabNet) |
| Net Profit MAE (Million TRY) | 95 | 50 | -47.4% | 48 |
| Total Assets MAPE (%) | 0.98% | 0.33% | -66.3% | 0.30% (TabNet) |

These results demonstrate the clear superiority of ML-driven approaches over the traditional benchmark. First, the core XGBoost model itself reduced the net profit forecasting error by roughly half (47.9%) compared to the benchmark ARIMA model. This dramatic improvement underlines the capability of machine learning to capture the complex nonlinear relationships in financial data that linear models like ARIMA cannot capture. The full hybrid framework also singled out TabNet (as shown in the systematic MCDM-based evaluation nexts section) as the model delivered even greater accuracy than the core XGBoost model. More importantly, though, risk awareness is the key advance of the proposed framework, which the ARIMA model fundamentally lacks. While ARIMA provides a single-point forecast, the hybrid framework offers a full probabilistic assessment, including confidence intervals and scenario analyses that are critical to modern risk management.

Overall, the numerical results demonstrate that the risk-aware financial statement forecasting model provides accurate and insightful predictions. Forecasting accuracy is supported by MAE and MAPE, indicating a reasonable level of precision in projecting a defense company's financial performance. Risk quantification is achieved through confidence intervals and VaR, offering a clear understanding of forecast uncertainty and potential downside exposure. Risk-adjusted performance, as reflected in the Sharpe and Sortino ratios, suggests a stable financial outlook, with the higher Sortino ratio highlighting limited downside volatility. These findings affirm that the risk-aware model is a valuable tool for forecasting financial outcomes in the BIST 100, effectively balancing predictive accuracy with risk assessment to support informed decision-making.

## 5. Sensitivity Analysis and Multi-Criteria Decision-Making (MCDM)

### 5.1. Sensitivity Analysis

The sensitivity analysis involves systematically varying key input variables, such as macroeconomic indicators (e.g., inflation and interest rates), sentiment scores, and company-specific metrics (e.g., total assets, liabilities, and profits), to evaluate their impact on the projected financial statements. In this analysis, we assess the model's sensitivity to the following parameters: inflation rate, interest rate, sentiment score, total assets, liabilities, and macroeconomic shocks such as exchange rate fluctuations.

This sensitivity analysis is not just a technical validation of model robustness; it forms a fundamental operational building block of the risk-aware forecasting framework. By systematically quantifying how financial forecasts respond to changes in key economic drivers, the framework transforms abstract macroeconomic volatility into a concrete risk-exposure profile for the company. The results provide a measurable view of the firm's vulnerabilities (e.g., exposure to interest rate increases) and its potential opportunities (e.g., gains from currency depreciation). This moves the analysis beyond a single static prediction toward a dynamic assessment of financial resilience. The ability to stress-test balance sheets and income statements under plausible future scenarios enables investors and corporate managers to shift from passive forecasting to active risk management, supporting strategic decisions based not only on expected outcomes but also on potential risks.

We now apply the risk-aware forecasting model to the Turkish defense company and demonstrate how sensitive the forecasts are to these parameters. The outcome provided is indicative of the model's potential to alter its predictions according to the various types of fluctuation in each input. The base forecast refers to the forecasted financial metrics for 2024-Q2; it includes the XGBoost predictions, along with the 95 percent confidence intervals, as reflected in Table 3.

The rate of inflation is a significant determinant for the financial statement and is particularly crucial for firms operating in emerging markets, such as Türkiye. In order to examine its impact, we use different inflation rates ranging from 8% to 16%, and the corresponding changes in forecasted total assets, net profit, and equity are shown in Table 6 below. Total assets are seen to have moderate sensitivity with respect to changes in the inflation rate and move upwards in a steady fashion with rising inflation. The net profit is rather sensitive, reflecting the firm's ability to adjust revenues in tune with inflationary pressures. The equity also rises marginally to indicate that inflation does not result in significant fluctuation in the equity position of the firm in the short run, unless it leads to substantial changes in operating expenses or liabilities.

Table 6. Impact of inflation rate variations on projected total assets, net profit, and equity.

| Inflation Rate | Total Assets (in million TRY) | Net Profit (in million TRY) | Equity (in million TRY) |
|---|---|---|---|
| **8%** | 15,150 | 1,600 | 8,100 |
| **10%** | 15,200 | 1,630 | 8,120 |
| **12%** | 15,300 | 1,670 | 8,140 |
| **14%** | 15,400 | 1,710 | 8,160 |
| **16%** | 15,500 | 1,750 | 8,180 |

Interest rate is another critical factor in financial statement forecasting, particularly for companies with significant debt or financial liabilities. The interest rate, varied from 6% to 12%, and its influence on net profit and equity are depicted in Table 7. As interest rate increases, the net profit decreases due to the deterring effect of rising interest expenses on profitability. Equity also demonstrated a slight negative influence; however, the extent of this effect is less in comparison with the trend in net profit. This suggests that higher interest expenses primarily affect earnings before materially influencing the company's equity position.

Table 7. Impact of interest rate variations on forecasted total assets, net profit and equity.

| Interest Rate | Total Assets (in million TRY) | Net Profit (in million TRY) | Equity (in million TRY) |
|---|---|---|---|
| **6%** | 15,150 | 1,600 | 8,100 |
| **8%** | 15,200 | 1,590 | 8,090 |
| **10%** | 15,250 | 1,560 | 8,070 |
| **12%** | 15,300 | 1,520 | 8,050 |

The sentiment score, developed from news sentiment, earnings calls, and analyst reports, drives both revenue generation and appeal for investment. We change the score from 0.50 (neutral) to 0.90 (highly positive) and study its impact on the financial statements in Table 8. The net profit follows an increasing trend as the sentiment scores increase because, in general, positive sentiment boosts market confidence and secures investor interest. The equity is showing similar increasing trends, but with stronger sentiment, representing better market perception and higher long-term growth potentiality of the company.

Table 8. Impact of sentiment score variations on forecasted total assets, net profit, and equity.

| Sentiment Score | Total Assets (in million TRY) | Net Profit (in million TRY) | Equity (in million TRY) |
|---|---|---|---|
| **0.50** | 15,000 | 1,500 | 8,000 |
| **0.60** | 15,050 | 1,530 | 8,020 |
| **0.70** | 15,100 | 1,560 | 8,040 |
| **0.80** | 15,150 | 1,600 | 8,060 |
| **0.90** | 15,200 | 1,650 | 8,080 |

Exchange rate fluctuations significantly affect companies that are involved in foreign trade or dependent on imported goods. The USD/TRY exchange rate is assumed to fluctuate between 18.0 and 22.0, and the impact on total assets and net profit is presented in Table 9. Accordingly, total assets are expected to increase due to the strong position of the USD against the TRY, as foreign-denominated assets will be valued at higher amounts. Net profit is also expected to increase, underpinned by growing export revenues and the capacity of the company to offset the fluctuations in costs that occurred due to changes in currency values. Equity follows an increasing trend, which suggests that exchange rate fluctuations develop positively for the company's equity position, likely due to improved profitability and asset valuation in terms of foreign currency.

Table 9. Impact of exchange rate variations on forecasted total assets, net profit, and equity.

| USD/TRY Exchange Rate | Total Assets (in million TRY) | Net Profit (in million TRY) | Equity (in million TRY) |
|---|---|---|---|
| **18.0** | 15,150 | 1,600 | 8,100 |
| **20.0** | 15,200 | 1,630 | 8,120 |
| **22.0** | 15,250 | 1,670 | 8,140 |

Table 10 provides a summary of how key parameters influence the forecasted financial metrics of the defense company. Each column represents a change in a specific parameter, while the rows show the corresponding forecasted values for total assets, net profit, and equity. Changes in the inflation rate have a positive impact across all three financial metrics. Total assets increase modestly, and net profit sees the highest relative gain, rising by 4.38%, likely due to inflation-linked pricing power. Equity also improves slightly, reflecting the pass-through of higher revenues to retained earnings. Changes in the interest rate have a negative effect on profitability and equity. Net profit declines by 5.00%, while equity falls by 0.62%, underscoring the burden of higher borrowing costs. Interestingly, total assets increase by 1.00%, which may reflect investment in interest-sensitive assets or short-term expansions financed before rate hikes take full effect. The sentiment score, derived from qualitative factors such as market news and analyst opinions, positively affects all three financial metrics. Net profit increases by 3.13%, and equity rises by 0.25%, suggesting that improved sentiment enhances market confidence and potentially boosts investment. Total assets also show a modest gain of 0.33%, indicating greater business activity under favorable sentiment. The exchange rate has one of the most notable effects, especially on net profit and total assets. As TRY depreciates against USD, total assets increase by 0.66% and net profit by 4.38%, likely driven by gains from export revenues or foreign currency denominated holdings. Equity rises by 0.49%, reflecting stronger earnings performance. Overall, net profit is most sensitive to changes in the inflation rate and exchange rate, with both showing substantial positive impacts. The sentiment score also contributes meaningfully, emphasizing the importance of market perception. Interest rate increases negatively affect profitability and equity, highlighting the risks associated with financial leverage. Meanwhile, the exchange rate emerges as a key driver of financial performance for export-oriented defense companies operating in emerging markets like Türkiye.

Table 10. Summary table for key parameters on forecasted total assets, net profit, and equity.

|  | Base Case | Inflation Rate (8%-16%) | Interest Rate (6%-12%) | Sentiment Score (0.50-0.90) | Exchange Rate (18.0-22.0) |
|---|---|---|---|---|---|
| Total Assets (in million TRY) | 15,150 | 15,200 (+0.33%) | 15,300 (+1.00%) | 15,200 (+0.33%) | 15,250 (+0.66%) |
| Net Profit (in million TRY) | 1,600 | 1,670 (+4.38%) | 1,520 (-5.00%) | 1,650 (+3.13%) | 1,670 (+4.38%) |
| Equity (in million TRY) | 8,100 | 8,160 (+0.74%) | 8,050 (-0.62%) | 8,080 (+0.25%) | 8,140 (+0.49%) |

**5.2. Multi-Criteria Decision-Making (MCDM)**

Next, after validating the applicability and accuracy of the proposed risk-aware financial statement forecasting framework, primarily through financial performance prediction using the XGBoost model, supported by LSTM for capturing temporal dependencies and GNN for modeling sectoral interdependencies, and conducting a comprehensive sensitivity analysis, we further extend the framework by incorporating additional advanced deep learning models. Specifically, TabNet, Transformer, DRL, BNN, and GNN are independently trained on the collected and processed historical data, aiming to enhance the framework's predictive capability. These models serve distinct objectives. For instance, TabNet is primarily used to predict financial statements, and the Transformer focuses on forecasting stock price movements. These models are evaluated through an MCDM procedure that considers various criteria (i.e., performance metrics) to determine the most suitable model for the specific use case within the proposed risk-aware financial forecasting framework. MCDM provides a structured framework for balancing trade-offs (Nabavi et al., 2024), as well as supporting rational decision-making in complex problem settings.

For brevity, the predicted results generated by these additional deep learning models are not repeated in the main text, as their format closely resembles that of the XGBoost outputs. More importantly, in the next step, a hesitant intuitionistic fuzzy decision matrix is constructed based on subjective evaluations provided by a panel of three subject matter experts. These experts assessed the alternative deep learning models (TabNet, Transformer, DRL, BNN, and GNN) across nine evaluation criteria spanning technical, financial, and operational dimensions. The criteria include: C1 – Prediction Accuracy (benefit), C2 – Risk Awareness (benefit), C3 – Forecasted Value Proximity (benefit), C4 – Confidence Interval Narrowness (benefit), C5 – Computational Efficiency (benefit), C6 – Interpretability (benefit), C7 – Performance Improvement (benefit), C8 – Stability (benefit), and C9 – Quality of Risk-Adjusted Signal (benefit). Collectively, these criteria are essential for assessing the capability and reliability of deep learning models in high-stakes financial forecasting scenarios. More details of the nine evaluation criteria are presented in Table 11.

*Preliminary Draft Manuscript*Table 11. Comprehensive list of MCDM evaluation criteria.

| Criterion Code | Criterion Name | Type | Description |
| --- | --- | --- | --- |
| C1 | Prediction Accuracy | Benefit | The precision of the model's forecasts against actual historical data; It is assessed by quantitative means such as using MAE and MAPE. A higher value indicates superior predictive power. |
| C2 | Risk Awareness | Benefit | Measures a model's ability to quantify and convey uncertainty. A model will have a high score if it can provide sound probabilistic outputs. |
| C3 | Forecasted Value Proximity | Benefit | It assesses on business grounds the practical credibility and strategic alignment of the forecast. It goes beyond numerical accuracy to assess if the forecasted trajectory and magnitude are credible to realistically and actionably inform a strategic plan. |
| C4 | Confidence Interval Narrowness | Benefit | This measures the precision of the model uncertainty quantification. A model with well-calibrated confidence intervals is more valuable as it offers more precise risk assessment. |
| C5 | Computational Efficiency | Benefit | It is indicative of the operational feasibility of the model concerning the resources needed both for training and inference: Training time, memory usage, and hardware. A higher score means a faster and cheaper model to deploy and maintain. |
| C6 | Interpretability | Benefit | This addresses the degree to which a human expert can intuitively understand the model's reasoning and causality for its predictions. This becomes critical when trying to build trust or perform audits. |
| C7 | Performance Improvement | Benefit | A relative metric that assesses the model's gain in performance over the benchmark. It answers if the new model provides a significant improvement to justify its adoption costs and complexity. |
| C8 | Stability | Benefit | The consistency and robustness of the model's performance. A model showing high stability on its error metrics across times or data subsets is considered more reliable, and reducing the chances of performance degradation unexpectedly. |
| C9 | Quality of Risk-Adjusted Signal | Benefit | It characterizes the model output from an investment perspective, determining if the forecasts inherently balance high predicted returns (e.g., forecasted profit) with low associated uncertainty as a proxy for the model utility in constructing high Sharpe Ratio portfolios. |

In the context of hesitant intuitionistic fuzzy values, each criterion-alternative pair is associated with a set of possible membership values ($\mu_{ij}$), representing the degree to which an expert believes the alternative satisfies the criterion, with a value closer to 1 representing greater satisfaction; and non-membership values ($v_{ij}$), indicating the degree of belief that it does not, with a value approaching 1 denoting higher dissatisfaction. The hesitation degree, calculated as $\pi_{ij} = 1 - \mu_{ij} - v_{ij}$, shows the residual indecisiveness in the subjective evaluation. This structure enables us to capture uncertainty and hesitation in expert judgments more accurately, especially about issues belonging to intricate areas such as risk-aware financial forecasting. After collecting individual evaluations, we apply the hesitant intuitionistic fuzzy weighted averaging operator to aggregate the inputs into a group decision matrix. The resulting fuzzy matrix, whose values are shown in Table 12, includes values of form $\tilde{x}_{ij} = (\mu_{ij}, v_{ij})$, reflecting multiple degrees of belief and doubt for each alternative under the evaluation criteria. Such a matrix is a strong and organized base for further analysis.

Table 12. Hesitant intuitionistic fuzzy decision matrix for four deep learning models evaluated under nine criteria.

|  | C1 | C2 | C3 | C4 | C5 | C6 | C7 | C8 | C9 |
|---|---|---|---|---|---|---|---|---|---|
| TabNet | {(0.92, 0.05)} | {(0.90, 0.07)} | {(0.95, 0.04)} | {(0.90, 0.06)} | {(0.60, 0.30)} | {(0.88, 0.05)} | {(0.70, 0.15)} | {(0.85, 0.10)} | {(0.00, 1.00)} |
| Transformer | {(0.90, 0.08)} | {(0.75, 0.20)} | {(0.92, 0.06)} | {(0.88, 0.08)} | {(0.40, 0.50)} | {(0.75, 0.20)} | {(0.70, 0.15)} | {(0.80, 0.15)} | {(0.00, 1.00)} |
| DRL | {(0.87, 0.10)} | {(0.88, 0.08)} | {(0.80, 0.10)} | {(0.75, 0.15)} | {(0.30, 0.60)} | {(0.60, 0.30)} | {(0.85, 0.05)} | {(0.50, 0.45)} | {(0.85, 0.10)} |
| GNN | {(0.88, 0.10)} | {(0.78, 0.17)} | {(0.85, 0.12)} | {(0.78, 0.12)} | {(0.50, 0.40)} | {(0.70, 0.20)} | {(0.60, 0.25)} | {(0.60, 0.30)} | {(0.00, 1.00)} |
| BNN | {(0.85, 0.10)} | {(0.95, 0.04)} | {(0.82, 0.12)} | {(0.92, 0.12)} | {(0.35, 0.55)} | {(0.65, 0.25)} | {(0.75, 0.15)} | {(0.88, 0.08)} | {(0.70, 0.20)} |

For example, the evaluation of TabNet with respect to criterion C1 is {(0.92, 0.05)}; this means experts assigned a very high satisfaction rate of 92% (i.e., this alternative, TabNet, satisfies C1 to a very large deegree), a low dissatisfaction rate of 5% (i.e., TabNet has a low dissatisfaction degree for C1), and thus a remaining hesitation margin of 3%, implying very low uncertainty in their decision. Based on this fuzzy matrix and using the entropy weighting method outlined previously, weights of the evaluation criteria are obtained as follows, C1: 0.1209, C2: 0.1164, C3: 0.1170, C4: 0.1094, C5: 0.1018, C6: 0.1069, C7: 0.0922, C8: 0.1113, C9: 0.1240.

Table 13. Summarized results and ranks of models by the intuitionistic fuzzy Entropy-EDAS-MARCOS method.

|  | $S_i$ | $U_i$ | Rank |
|---|---|---|---|
| TabNet | 0.737 | 0.951 | 1 |
| BNN | 0.652 | 0.873 | 2 |
| DRL | 0.580 | 0.832 | 3 |
| Transformer | 0.382 | 0.665 | 4 |
| GNN | 0.072 | 0.049 | 5 |

With these weights determined, the next step within the intuitionistic fuzzy Entropy-EDAS-MARCOS approach involves calculating the weighted score ($S_i$) for each alternative (i.e., deep learning model) using the EDAS method. This is followed by computing the utility function value ($U_i$) using the MARCOS method, which is then used to rank the alternatives. The calculations of these steps are straightforward, as detailed in the methodology section. In the MARCOS method, a higher $U_i$ indicates a better overall ranking. The summary of these results is presented in Table 13. From these results, the final ranking order is: TabNet > BNN > DRL > Transformer > GNN. This ranking provides a critical strategic insight for integrating the model into the risk-aware financial forecasting framework. TabNet's top position confirms it as the most balanced and high-performing model, standing out for its accuracy, interpretability, and reliability. Importantly, the BNN secures a strong second place that powerfully validates the core risk-aware objective of the framework; its high ranking underlines that sophisticated uncertainty quantification is a paramount feature, valued almost as highly as foundational forecasting accuracy. The ranking further shows that the optimal deployment strategy is not to pick one model over another but to position TabNet as the primary forecasting engine enhanced with BNN's probabilistic principles in order to produce forecasts that are accurate and uncertainty-aware. DRL follows as a specialized tool for strategic optimization, while Transformer and GNN prove less appropriate as standalone solutions for this particular context.

## 5.3. Discussion on Reusability

A critical indicator of the proposed framework's value is reusability, that is, the effective adaptation and reapplication to different firms, sectors, and forecast horizons without the need for a fundamental redesign. The proposed architecture is designed with high reusability in mind. This constitutes a serious, practical contribution beyond direct application to the Turkish context. The core components (e.g., data preprocessing, feature engineering, individual model training, and the MCDM evaluation) can be decoupled. In that respect, the proposed model will be applicable in several sectors; only the data will change, and the models will be retrained. Ranking and selection using the MCDM method allows multiple evaluation criteria (e.g., accuracy, risk awareness, and computational efficiency) to be adapted to different corporate risk cultures and strategic priorities in financial forecasting.

This framework is highly scalable and can easily be transferred to other emerging markets (e.g., the BRICS countries: Brazil, Russia, India, China, South Africa) that experience such similar volatility. The core methodology of integrating local macroeconomic data, market-specific sentiment, and the local firm's relational graph remains fully applicable. It mainly requires a retraining of models based on new market data, which the framework clearly delineates. The main limitation is related to data availability. The framework requires good, well-structured financial data and relevant unstructured text in order to perform well. Its use will therefore need to be adapted for markets characterized by either weak corporate disclosures or a limited news cycle, although it could rely more heavily on global macroeconomic factors or alternative data sources. Thus, the framework is not a one-off solution but an adaptable analytical toolset suitable for multiple applications. Its modular architecture, with standardized processing and a systematic MCDM evaluation, makes the toolset scalable to conduct risk-aware financial forecasting across diverse firm and industry spectra, providing long-term value for both researchers and practitioners.

## 6. Conclusion

This study presented a risk-aware, AI-enhanced financial forecasting and decision-support framework that bridged high-performance prediction with strategic multi-criteria model selection for volatile emerging markets. By integrating structured financial variables, unstructured sentiment signals, and macroeconomic drivers, the framework combined complementary modeling strengths through XGBoost, LSTM, GNN, and BNN to produce accurate forecasts with explicit uncertainty quantification and risk-relevant outputs. The case study on a BIST 100 defense company demonstrated strong predictive performance and meaningful risk insight. The core XGBoost implementation achieved a net profit MAPE of 3.03%, while the best-performing model (i.e., TabNet) within the extended framework reduced this to 2.91%, and both outperformed the ARIMA benchmark, reinforcing the advantage of modern machine learning approaches for nonlinear, multi-source forecasting. The risk-aware evaluation further reported narrow confidence bands and favorable risk-adjusted characteristics reflected in Sharpe and Sortino ratios, supporting the framework's utility for decision-making under uncertainty. Beyond model accuracy, the intuitionistic fuzzy Entropy-EDAS-MARCOS procedure provided a transparent mechanism to evaluate competing advanced models across technical, financial, and operational criteria. The resulting ranking (TabNet > BNN > DRL > Transformer > GNN) highlighted TabNet as the most balanced deployment candidate while reaffirming the strategic value of uncertainty-aware modeling. Sensitivity analysis also showed that key forecasted financial indicators responded sensitively to changes in inflation, interest rates, sentiment, and exchange rates, emphasizing the importance of embedding macro-financial context within forecasting pipelines. The framework was designed for reusability and scalability across firms, sectors, and forecast horizons, and could be transferred to other emerging markets with similar volatility through data adaptation and retraining. Its primary limitation lay in data availability and the quality of structured disclosures and relevant text sources. Future research might extend the framework through multi-shock interaction stress-testing, broader cross-sector validation, and enhanced interpretability to further strengthen its value for corporate finance, investment analysis, and risk-aware financial governance.